\documentclass[conference]{IEEEtran}
\IEEEoverridecommandlockouts
\pdfoutput=1
\usepackage{cite}
\usepackage{amsmath,amssymb,amsfonts}
\usepackage{algorithmic}
\usepackage{graphicx}
\usepackage{textcomp}
\usepackage{xcolor}
\def\BibTeX{{\rm B\kern-.05em{\sc i\kern-.025em b}\kern-.08em
    T\kern-.1667em\lower.7ex\hbox{E}\kern-.125emX}}
\usepackage{multirow}
\usepackage{url,xcolor}
\usepackage{float}
\usepackage[hidelinks]{hyperref}

\newboolean{showcomments}
\setboolean{showcomments}{true}
\ifthenelse{\boolean{showcomments}}
{ \newcommand{\mynote}[3]{
    \fbox{\bfseries\sffamily\scriptsize#1}
    {\small$\blacktriangleright$\textsf{\emph{\color{#3}{#2}}}$\blacktriangleleft$}}}
{ \newcommand{\mynote}[3]{}}

\begin{document}

\title{The Influence of Reward on the Social Valence of Interactions\\
\thanks{This work was supported by national funds through Funda\c c\~ao para a Ci\^encia e a Tecnologia (FCT) with references SFRH/BD/144798/2019, SFRH/BD/143460/2019, and UIDB/50021/2020.}
}

\author{
	\IEEEauthorblockN{Tom\'as Alves}
	\IEEEauthorblockA{
		\textit{INESC-ID \& Instituto Superior T\'ecnico,}\\
		\textit{University of Lisbon, Portugal}\\
		tomas.alves@tecnico.ulisboa.pt
	}
	\\[0.3cm]
	\IEEEauthorblockN{Jo\~ao Dias}
	\IEEEauthorblockA{
		\textit{INESC-ID \& Faculty of Science and Technology,}\\
		\textit{University of Algarve, Portugal}\\
		joao.dias@gaips.inesc-id.pt
	}
	\\
	\and
	\IEEEauthorblockN{Samuel Gomes}
	\IEEEauthorblockA{
		\textit{INESC-ID \& Instituto Superior T\'ecnico,}\\
		\textit{University of Lisbon, Portugal}\\
		samuel.gomes@tecnico.ulisboa.pt
	}
	\\[0.3cm]
	\IEEEauthorblockN{Carlos Martinho}
	\IEEEauthorblockA{
		\textit{INESC-ID \& Instituto Superior T\'ecnico,}\\
		\textit{University of Lisbon, Portugal}\\
		carlos.martinho@tecnico.ulisboa.pt
	}
}

\maketitle

\begin{abstract}
  Throughout the years, social norms have been promoted as an informal enforcement mechanism for achieving beneficial collective outcomes.
  %Although there are several methods to foster interactions with different social valence values such as framing the context dilemma or setting in-game rules, to the best of our knowledge there is a lack of research regarding the use of incentives.
  Among the most used methods to foster interactions, framing the context of a situation or setting in-game rules have shown strong results as mediators on how an individual interacts with their peers.
  Nevertheless, we found that there is a lack of research regarding the use of incentives such as scores to promote social interactions differing in valence.
  Weighing how incentives influence in-game behavior, we propose the use of rewards to promote interactions varying in valence, i.e. positive or negative, in a two-player scenario.
  To do so, we defined social valence as a continuous scale with two poles represented by \textit{Complicate} and \textit{Help}.
  Then, we performed user tests where participants where asked to play a game with two reward-based systems to test on whether the scoring system influenced the social interaction valence.
  The results indicate that the developed reward-based systems were able to foster interactions diverging in social valence scores, providing insights on how factors such as incentives overlap individual's established social norms.
  These findings empower game developers and designers with a low-cost and effective policy tool that is able to promote in-game behavior changes.
\end{abstract}


\section{Introduction}

Recent research has focused on how social norms have an impact on behavior in a wide variety of contexts~\cite{Bicchieri2018,nicolardi2020pain,young2015evolution}.
In particular, social norms have been promoted as an informal enforcement mechanism for achieving beneficial collective outcomes~\cite{nyborg2016social}.
Our work focuses on the social valence of interactions, which addresses the valence, i.e. positive or negative, of how an individual tries to interact with their peers.
In particular, we define interdependent social valence as a continuous scale with two poles represented by \textit{Complicate} and \textit{Help}.
There are several factors that may have an impact on the valence of social interactions.
Kokkinakis et al.~\cite{kokkinakis2016s} found that, in the context of a multiplayer online battle arena game, usernames containing either player age or highly anti-social words are correlated with the valences of player interactions within the game.
On the research field of social dilemmas, Martinez et al.~\cite{martinez2011behavioural} studied whether regret and disappointment could be distinguished on the basis of how these emotions manifest themselves in social-dilemma situations.
Their results showed that while regret increases pro-social behavior, disappointment provokes the opposite effect.
In particular, regret led to more generous offers whereas disappointment led to less generous offers while playing social bargaining games.

% With the rise of ubiquitous technologies, researchers have been able to employ new and innovative methods such as serious games to promote behavioral modification with the expansion of communication and media technologies.
Besides social dilemmas, researchers have examined the development of interaction styles diverging in social valence in general-purpose serious games.
For example, Vegt et al.~\cite{Vegt2016} analysed how game rules change interdependent behavior.
Results showed that different game rules could generate distinct reported player experiences and observable distinct player behaviors that could be further discriminated into four patterns: expected patterns of helping and ignoring, and unexpected patterns of agreeing and obstructing.
%Nevertheless, the valence of social interactions may be affected by other factors such as incentives.
Besides game rules, incentives (e.g. personal satisfaction, rewards, praise, or recognition) have been shown to be a cornerstone in behavioral economics~\cite{Zandstra2013}, acting as behavioral moderators~\cite{Baranowski1997}.
Therefore, game designers should pay attention to the game elements, e.g. rules, avatars, levels, or incentives, they employ to provide in-game assessment of player performance.
% since they are an important factor to promote behavior and attitude changes~\cite{Hagood2016}.}
To the best of our knowledge, there is a lack of research on the application of rewards, notably scoring systems, as a means to promote in-game behavioral change, in particular interactions with different social valence.

%Weighting how social norms provide insights regarding factors that influence the valence of an individual's specific actions,
In the light of this, our research goal is to \textbf{study how rewards have an effect on the social valence of interactions in a two-player scenario}.
%, which address the valence of what a player tries to do to the other player while they play a game.
As an evaluation platform, we used a word-matching game named Message Across\footnote{The code for the game is available online, hosted in the platform \textit{GitHub}: \url{https://github.com/SamGomes/message-across} (as verified at \today).}, in which we developed two versions of its reward attribution system, one aimed at promoting competition and the other at promoting mutual help. 
Notably, both versions keep the same mechanics except for the scoring system, which is unknown to both participants.
This design decision induces the need of both players to decode which is the best strategy to employ for each version depending on the scoring system.
% Message Across is a two-player \textit{in loco} game with different reward systems to foster four different in-game behaviors: competition, self-improvement, mutual help, and extreme altruism.

The paper is organized as follows: In Section~\ref{sec:relatedWork}, we present studies that address the social valence of interactions on interdependent scenarios.
Section~\ref{sec:model} follows, with the conceptualization of our model to study interdependent social interaction valence.
Then, Section~\ref{sec:method} presents the method used to collect data for our work, and Section~\ref{sec:results} provides results and discusses our contributions.
Finally, we conclude our work with a discussion on future directions, tackling the limitations of our current approach.

\section{Related Work}
\label{sec:relatedWork}

Behavior change has been an emerging topic in the last decade, covering a wide range of topics from identifying and targeting bystander behavior and cultural issues in cyberbullying~\cite{Ferreira2016,Ferreira2019} to dealing with climate change issues~\cite{Lee2015,Ouariachi2019}.
In particular, serious games have been used as an effective channel delivering behavior change~\cite{pellegrini2009role,Mandryk2013,Chow2020} and fostering activities that lead to learning~\cite{Ryan1991,pellegrini2009role,lieberman2012designing}, such as in the health~\cite{Mandryk2017,joyner2017fit,wang2017story} or social science~\cite{Schonbrodt2011} behavior fields.
Nevertheless, the approach to promote behaviors has a strong impact on the social valence of interactions.
For instance, while video games may motivate prosocial behaviors by having prosocial content~\cite{Gentile2009,Greitemeyer2010,greitemeyer2013exposure}, they may also lead to negative behaviors, e.g. increased aggression right after playing violent games~\cite{Anderson2001,Anderson2004}.

Research on social dilemmas (e.g. prisoner's dilemma and dictator games) has focused on identifying factors that can explain and improve cooperative behavior~\cite{ellingsen2012social,goerg2019framing}.
Böhm and Theelen~\cite{bohm2016outcome} addressed the influence of outcome valence -- whether individuals will have a positive or negative outcome, irrespective of their behavior~\cite{bohm2016outcome} -- and externality valence -- whether an individual's transfer to the public good will have a positive or negative effect on others~\cite{bohm2016outcome} -- framing on players' willingness to cooperate in a repeated public good game.
The authors found that individuals' social value orientation is related to cooperative behavior.
In particular, there was an increased cooperation when individuals faced a situation where they needed to decrease a collective loss by transferring tokens that caused positive externalities on others.

On the same line of research, Farrow et al.~\cite{Farrow2018} studied whether valence framing impacts the effectiveness of a social norm intervention on prosocial behavior.
The authors used two variations of the dictator game~\cite{balafoutas2017nature} that are structurally equivalent in the range of payoff outcomes and their respective theoretical predictions.
While the positive valence version shows a worker that has \$1.00 and is given the opportunity to allocate this amount between himself and another worker with whom he is randomly paired, the negative version depicts a worker that is randomly paired with another worker who is endowed with \$1.00, and the first worker is given the opportunity to make the same allocation decision.
The task of the dictator is to distribute an amount of money between himself and a worker with whom he is paired under two different social norm interventions: the dictator is told that most of the dictators in previous sessions transferred to the worker an average of either \$0.30 or \$0.50.
Results showed that a social norm intervention can have a significant positive impact on allocation decisions in the positive frame, although this is only true for participants that were told that dictators in previous sessions transferred an average of \$0.50 to the other worker.
Regarding the negative frame, there was also a significant average treatment effect of framing with respect to the social norm interventions, since the positive frame elicited an allocation of \$0.310 and the negative frame \$0.393.
Follow-up regression analysis suggests that there may exist important dissimilarities on how participants decided to allocate money across frames.
These findings show that it is possible to model behavior based on how researchers frame the decision context.

Another approach to foster behaviors is by defining sets of in-game rules.
In order to study how player behavior that relates to team performance can be affected by game rules, Vegt et al.~\cite{Vegt2016} developed two versions of a multiplayer game, varying in rule-sets, designed to elicit player strategies of interdependence and teamwork.
While interaction rules were expected to evoke a strong feeling of interdependence, goal-driven rules were expected to elicit competition in one game variant, and cooperation in the other.
These two rules-set framework defines four interdependent game behaviors between goal-driven and interaction-rules: \textit{dependent competition}, \textit{independent competition}, \textit{dependent cooperation}, and \textit{independent cooperation}.
Results show that competition and cooperation goal-driven rules led players to experience competition and cooperation, respectively.
Regarding interaction rules, they mainly stimulated dependent competitive behavior, e.g. obstructing each other, suggesting that the interaction rules may have overruled the effect of goal-driven rules in the competitive game.

Finally, score systems provide a flexible approach to feedback, and can easily imply direct competition due to easy comparisons between players~\cite{richter2015studying}, thus helping to differentiate social valence levels regarding interactions.
Cress and Martin~\cite{cress2006knowledge} studied the effects of rewards and punishments in a public goods game.
The authors concluded that both rewards and punishments contributed to higher public player investments.
%In particular, results suggest that rewarding contributions with a cost compensating bonus can be an effective solution not only at the individual level, but also at the group level.
Furthermore, when players did not know each other, the effect was significantly more predominant when using punishment.
Given these prospects, we believe that studying how to influence the social-valence of in-game interactions is relevant to advance knowledge in this field.

\section{Towards an Interdependent Social Valence Model}
\label{sec:model}

\begin{figure}[!htbp]
	\centering
	\includegraphics[width=1.0\columnwidth]{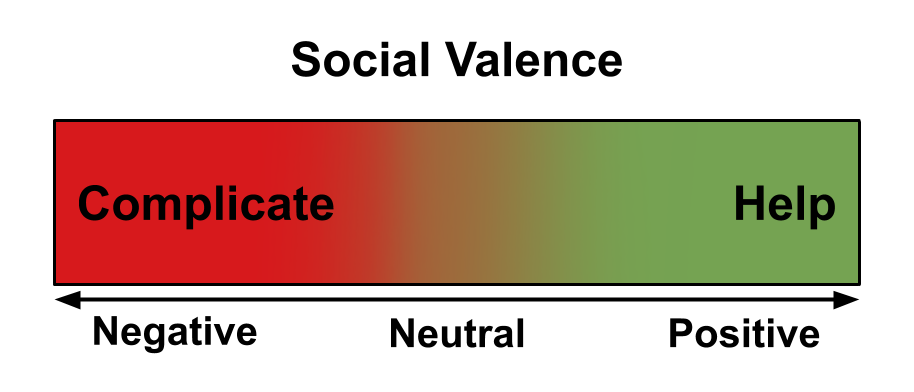}
	\caption[Social Valence Scale.]{Social Valence Scale.}
	\label{fig:socialValenceScale}
\end{figure}

Based on the aforementioned research, we focus on the nature of one's intentions towards other players by formulating social valence as a composition of a ``social" component -- there is interaction between peers -- and a ``valence" component -- whether my intention is positive or negative towards the other player.
Notably, we address it with a continuous scale with two poles: \textit{Complicate} and \textit{Help} (see Figure~\ref{fig:socialValenceScale}).
% In terms of interaction style, the scale measures the valence between cooperation (helping) and competition (complicating).
Firstly, we believe the positive pole is aligned with prosocial behavior, representing an intent or behavior to benefit others or the society in general.
However, this does not mean that a prosocial behavior necessarily implies that an individual has to sacrifice his own goals.
In our model, the positive pole represents an interaction style of \textit{Mutual Help}, leading an individual to equally focus on both themself and the others while performing their task.
Additionally, we argue that a positive valence is often correlated with behavior that promotes social harmony.
For instance, collectivistic cultures generally have cultural norms that promotes social harmony~\cite{callister1997japanese}.
In the context of a game, this can be modeled as player focusing on advancing their task while minimizing the difference between their task progression compared to their peers.
%Firstly, we believe that the positive pole is connected to the interaction style \textit{Mutual Help}, as this style of interaction leads an individual to focus on both themself and the other player while performing their task.
%In particular, a mutual helper will concurrently try to achieve their own goals and, whenever possible, help others to succeed.
%\textcolor{blue}{Therefore, we believe that an individual will focus on minimizing the difference between their task progression compared to their peers.
%In the light of this,} \textit{Mutual Help} is associated to the positive social valence of interactions.
Secondly, we propose that the negative pole is connected to the interaction style \textit{Competition}.
In fact, although being of negative valence, competitive interactions and competition fostering game elements are known to play an imperative role on improving players' enjoyment and learning~\cite{cagiltay2015effect,consalvo2011using,vorderer2003explaining}.
Thus, \textit{Competition} assumes that the player's goal should not be bound to improve themself, but instead to become better than other players (i.e. improve their rank in the game).
To reach this goal, we believe that a competitive player can embrace two strategies: either the player progresses further in the game when there is no meaningful interaction, or they complicate and sabotage the other players' task progression when there is a meaningful interaction.
With this in mind, a competitive individual will focus on maximizing the difference between task progression compared to their peers.

\section{Method}
\label{sec:method}

As we have previously mentioned, our goal is to understand whether rewards have an effect on the social valence of interactions in a two-player scenario.
In particular, we evaluate the effectiveness of score attribution systems in promoting different social valence values.
Therefore, our first research question stands as:
\begin{itemize}
    \item \textit{How does reward affect the social valence of interactions in a two-player scenario?}
\end{itemize}

In addition, we expect that fostering interactions with contrasting social valence values will have an effect on the difference between task progression among peers.
In other words, the difference between scores at the end of the experience will depend on the scoring system.
Hence, our second research question is:
\begin{itemize}
    \item \textit{How does reward affect the difference between final scores in a two-player scenario?}
\end{itemize}

The next sections cover the methodology used to collect data to address our research questions.

\subsection{Participants}

Subjects were recruited in pairs through standard convenience sampling procedures including direct contact and through word of mouth.
Subjects included any local interested in participating if they were at least 18 years old.
Each participant was asked to sign a consent form.
There were no potential risks and no anticipated benefits to individual participants. 
We conducted a total of 37 tests.
After data analysis, four tests did not meet quality criteria, e.g in-game data not recorded or questionnaires with missing answers.
Thus, our final data set comprises 33 tests, a total of 66 participants (37 males, 29 females) between 18 and 40 years old \((M = 23.12; SD = 4.09)\).

\begin{table*}[!htbp]
    \caption{Rewards attributed to each player based on the action a player performs, whether the letter is useful for them, how is their task progression compared to the other player, and the scoring system.}
    \centering
    \label{tbl:scoringSystems}
    \begin{tabular}{|c|c|c|c|c|c|c|c|}
    \hline
    \multirow{2}{*}{\textbf{\textit{Action}}}  & \multicolumn{2}{|c|}{\textbf{\textit{Usefulness}}} & \multirow{2}{*}{\textbf{\textit{Task Progression}}} & \multicolumn{2}{|c|}{\textbf{\textit{Mutual Help}}} & \multicolumn{2}{|c|}{\textbf{\textit{Competition}}} \\\cline{2-3}\cline{5-8}
    & \textbf{Me} & \textbf{Other} & & \textbf{Me} & \textbf{Other} & \textbf{Me} & \textbf{Other} \\ \cline{1-8}
    \multirow{12}{*}{\textit{GIVE}} & \multirow{3}{*}{\textit{YES}} & \multirow{3}{*}{\textit{YES}} & \textit{I am ahead} & 10 & \multirow{3}{*}{0} & \multirow{3}{*}{0} & \multirow{3}{*}{0} \\ \cline{4-5}
    & & & \textit{We are equal} & 10 & & & \\ \cline{4-5}
    & & & \textit{I am behind} & 0 & & & \\ \cline{2-8}
    &\multirow{3}{*}{\textit{NO}} & \multirow{3}{*}{\textit{YES}} & \textit{I am ahead} & 10 & \multirow{3}{*}{0} & \multirow{3}{*}{0} & \multirow{3}{*}{0} \\ \cline{4-5}
    & & & \textit{We are equal} & 10 & & & \\ \cline{4-5}
    & & & \textit{I am behind} & 0 & & & \\ \cline{2-8}
    & \multirow{3}{*}{\textit{YES}} & \multirow{3}{*}{\textit{NO}} & \textit{I am ahead} & \multirow{3}{*}{0} & \multirow{3}{*}{0} & \multirow{3}{*}{0} & \multirow{3}{*}{0} \\ \cline{4-4}
    & & & \textit{We are equal} & & & & \\ \cline{4-4}
    & & & \textit{I am behind} & & & & \\ \cline{2-8}
    &\multirow{3}{*}{\textit{NO}} & \multirow{3}{*}{\textit{NO}} & \textit{I am ahead} & \multirow{3}{*}{0} & \multirow{3}{*}{0} & \multirow{3}{*}{0} & \multirow{3}{*}{0} \\ \cline{4-4}
    & & & \textit{We are equal} & & & & \\ \cline{4-4}
    & & & \textit{I am behind} & & & & \\ \cline{1-8}
    \multirow{12}{*}{\textit{TAKE}}&\multirow{3}{*}{\textit{YES}} & \multirow{3}{*}{\textit{YES}} & \textit{I am ahead} & 0 & \multirow{3}{*}{0} & \multirow{3}{*}{5} & \multirow{3}{*}{-5} \\  \cline{4-5}
    & & & \textit{We are equal} & 10 & & & \\  \cline{4-5}
    & & & \textit{I am behind} & 10 & & & \\  \cline{2-8}
    &\multirow{3}{*}{\textit{NO}} & \multirow{3}{*}{\textit{YES}} & \textit{I am ahead} & \multirow{3}{*}{0} & \multirow{3}{*}{0} & \multirow{3}{*}{0} & \multirow{3}{*}{-5} \\  \cline{4-4}
    & & & \textit{We are equal} & & & & \\  \cline{4-4}
    & & & \textit{I am behind} & & & & \\  \cline{2-8}
    &\multirow{3}{*}{\textit{YES}} & \multirow{3}{*}{\textit{NO}} & \textit{I am ahead} & \multirow{3}{*}{10} & \multirow{3}{*}{0} & \multirow{3}{*}{5} & \multirow{3}{*}{0} \\  \cline{4-4}
    & & & \textit{We are equal} & & & & \\  \cline{4-4}
    & & & \textit{I am behind} & & & & \\  \cline{2-8}
    &\multirow{3}{*}{\textit{NO}} & \multirow{3}{*}{\textit{NO}} & \textit{I am ahead} & \multirow{3}{*}{0} & \multirow{3}{*}{0} & \multirow{3}{*}{0} & \multirow{3}{*}{0}  \\  \cline{4-4}
    & & & \textit{We are equal} & & & & \\  \cline{4-4}
    & & & \textit{I am behind} & & & & \\
    \hline
    \end{tabular}
\end{table*}

\subsection{Apparatus}

\begin{figure}[!htbp]
	\centering
	\includegraphics[width=0.95\columnwidth]{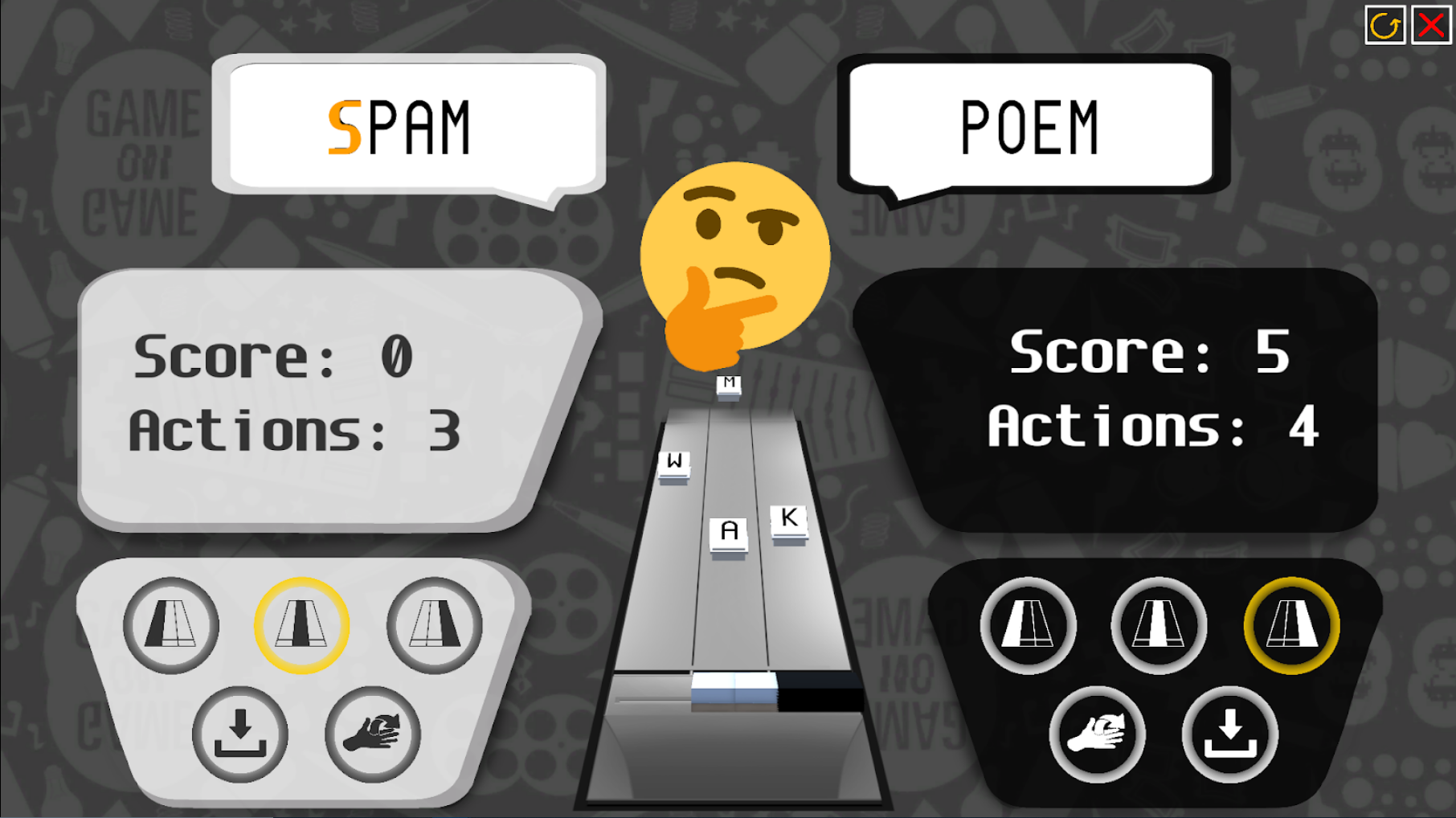}
	\caption[Screenshot of MessageAcross.]{Screenshot of MessageAcross.}
	\label{fig:messageacross}
\end{figure}

As we previously mentioned, we used the Message Across game (Figure~\ref{fig:messageacross}) to develop different versions of the score attribution system.
Message Across is a two-player game where each individual has a four-letter word to complete for each level, placed on the top of the screen.
We decided that words always share two letters in order to promote interactions between players, since both of them need to interact with those specific letters.
Letters are presented on three different lanes, and players can move their cursor to a lane and interact with a letter.
If both players are in the same lane, only the first player to select an action is able to perform it, i.e. if both players are in the same lane and player A presses and holds before player B an action button, the action of player A is considered while player B's is ignored.
There are two different ways to interact with a letter: a player can decide to give a letter to the other player or to take a letter for themself.
Notably, a player only has four actions to perform per level.
This design decision is based on the fact that, by having a limited number of actions per level, players will consider the cost of opportunity of keeping or giving a letter and follow meaningful strategies, while still being able to finish their four-letter word.
When both players run out of actions, the level ends.
The reward system adjusts the score that is given to a player based on their action (\textit{give} or \textit{take}) according to the induced interaction style.

The scoring systems are presented in Table~\ref{tbl:scoringSystems}.
There are some rewards that are similar to both scoring systems such as when a player \textit{takes} a letter that they do not need or \textit{gives} a letter that their peers do not need, neither player receives reward.
On a singular level, the \textit{Competition} version focuses on promoting a behavior that increases the difference between the task completion of the player and their peers.
As such, players have three different strategies to follow.
Firstly, they can advance their task progression and punish the other at the same time.
Thus, when both players need the same letter, this version rewards players who take letters (their score increases by five points) and punishes the other players (the other players' score decreases by five points).
Secondly, they can progress on their task by \textit{taking} letters that the other player does not want, where their score increases by five points.
Thirdly, players can just punish their peers by taking a letter that the other needs, while they do not need it.
This leads to a decrease in five points in the score of the other player.
Moreover, \textit{give} actions offer no reward to the player, as they would not contribute to a maximization of the difference in task completion between players.

In contrast, the \textit{Mutual Help} version aims at minimizing the gap between the players' task progressions.
Therefore, the most important interaction scenarios are when a player tries to interact with a letter that the other player needs.
When a player has less acquired letters than the other and \textit{gives} a letter that either they both need or only the other player needs, neither player receives reward, as they would not be minimizing the difference between the number of letters they have already collected.
The same happens if a player has more letters than the other and \textit{takes} a letter that they both need.
In this case, \textit{giving} the letter to the other player is the best option since it would decrease the difference between the number of acquired letters.
While there is a similarity with the \textit{Competition} version where players gain score by \textit{taking} letters that the other player does not want, the \textit{Mutual Help} version specifically promotes \textit{taking} letters that both players need when the player that performed the action has the same number or less acquired letters than the other participant, as it decreases difference between task progressions.

%In particular, each version had the following score systems:
%\begin{itemize}
	%\item The \textbf{Competitive} system rewards players who take letters (their scores increase by five points), punishing the other players if the taken letters are shared between players (the other players' scores decrease by five points).
	%\item The \textbf{Mutual Help} system rewards with ten points players: (A) who give letters when they were ahead or take letters when they were behind; and (B) who take letters that are not shared.
%\end{itemize}

The experimental setup included additional material, namely (i) a tutorial version of Message Across which does not grant reward to any player for a \textit{give} or \textit{take} action, (ii) a questionnaire with one item in a seven-point Likert Scale translated to the local language to address the social valence of the interactions the player performed regarding the other player -- ``What did you try to do to the other player?" from ``Complicate" (-3) to ``Help" (3) --, (iii) a questionnaire to assess the preferred game version, (iv) a computer to run the game, (v) another computer to allow the participant to fill in the questionnaires, (vi) a touchscreen monitor to play the game, and (vii) a GoPro video camera put on a tripod and positioned approximately 50 cm in front of the touch screen for observation of player movement and in-game activity.

\subsection{Procedure}

\begin{figure}[!htbp]
	\centering
	\includegraphics[width=0.95\columnwidth]{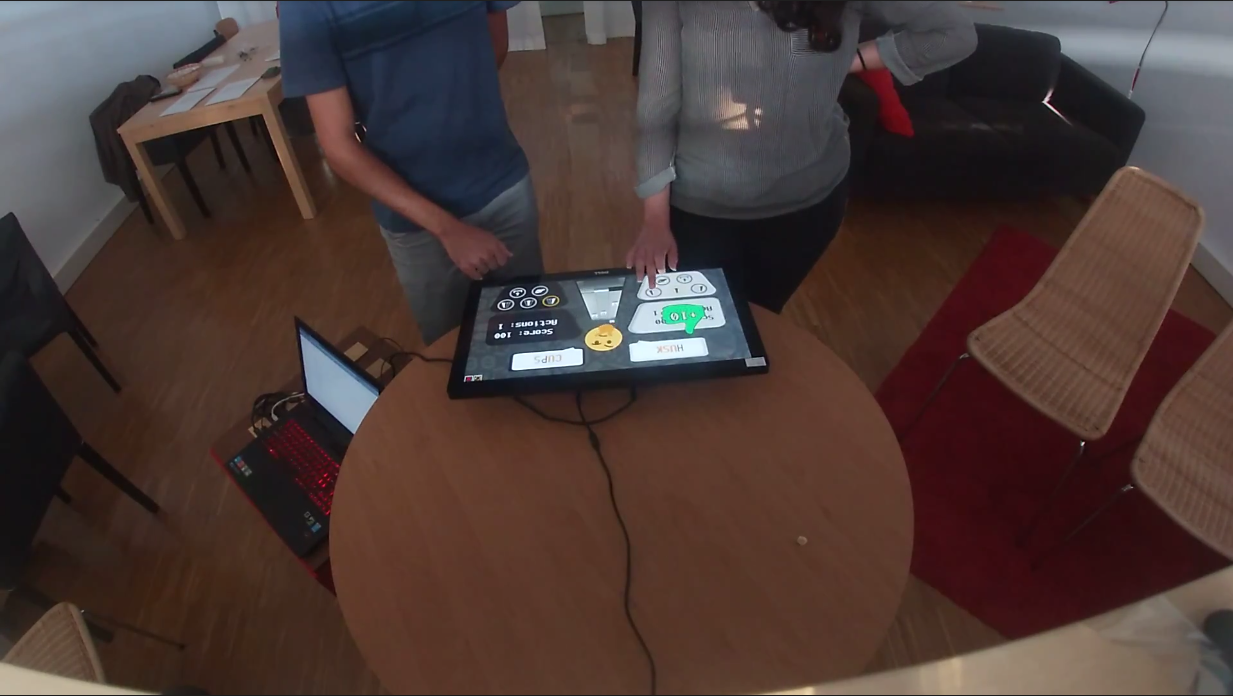}
	\caption[Experimental setup.]{Experimental setup.}
	\label{fig:setup}
\end{figure}

Before the experiment, participants were informed about the experience and invited to sign a compulsory consent form.
They were also informed that they could stop the experiment at any time.
After receiving consent, both participants were asked to be next to the touchscreen (as seen in Figure~\ref{fig:setup}) and received a tutorial regarding in-game mechanics and possible actions to perform.
Additionally, we allowed participants to play up to seven rewardless levels in the tutorial version in order to support the development of minimal skills to play the game.

When both participants felt comfortable with the game mechanics, they played the two game versions in random order without knowing which version they were playing at each moment.
Notably, neither player knew how the scoring system rewarded interactions, which led the participants to try to understand it while playing the game.
Each gaming session lasted for seven levels.
After each gaming session, participants were asked to fill in a scale addressing social valence while playing that version.
At the end of the experiment, participants were invited to fill a questionnaire to choose their preferred version of the game and received their compensation.

\section{Results}
\label{sec:results}

\begin{figure}[!htbp]
	\centering
	\includegraphics[width=0.95\columnwidth]{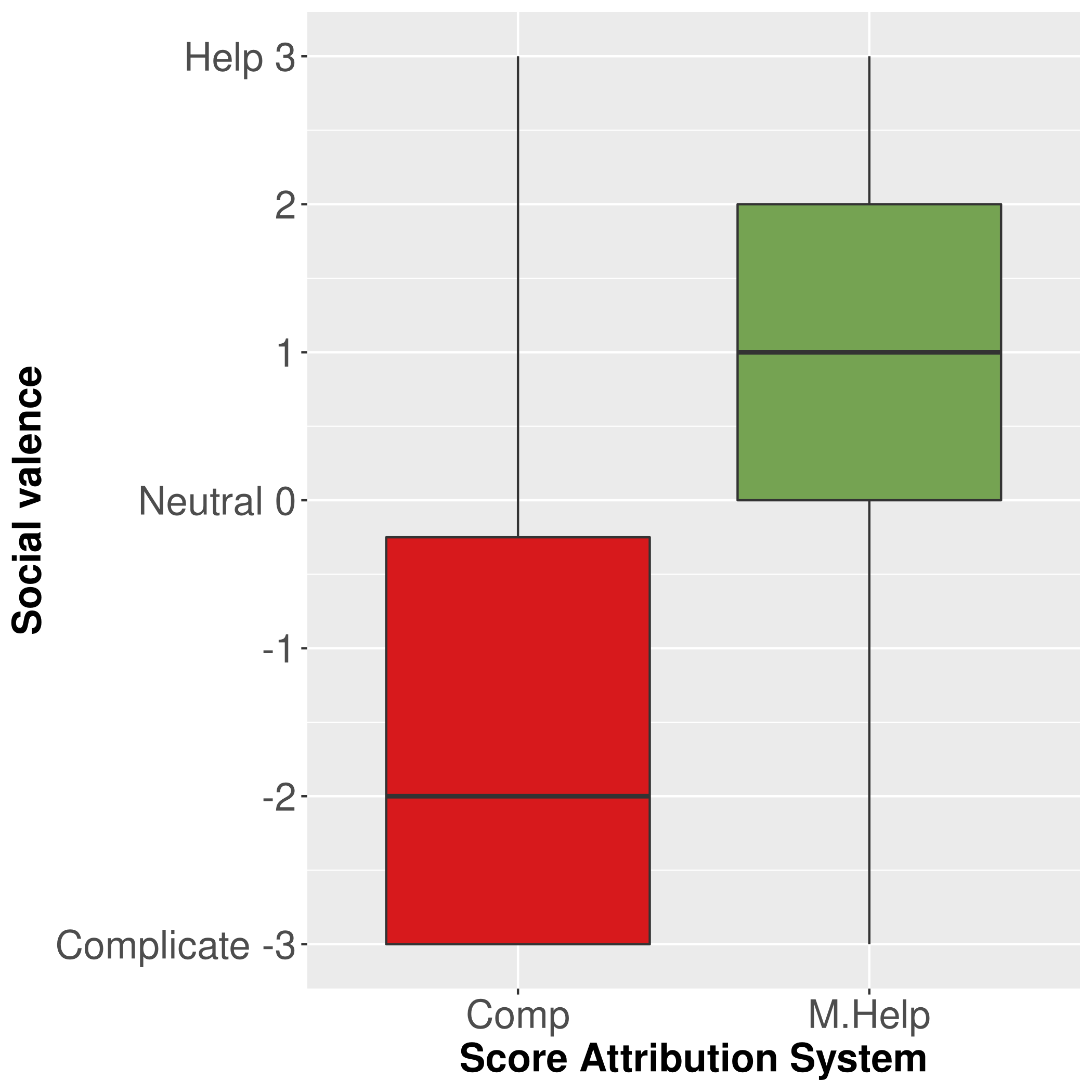}
	\caption{Distribution of average interaction intention values by score system.}
	\label{fig:socialValence}
\end{figure}

Our research questions address one independent categorical variable, the score attribution system (with two levels, \textit{Competition} and \textit{Mutual Help}) and two dependent variables -- the social valence, bounded between \textit{Complicate} and \textit{Help}, and the difference between the final scores of both players.
As we mentioned, the social valence, representing what the player reported as the dominant intention while playing each game version, was measured at the end of each played version through the question ``What did I try to do to the other player while playing this version?''.
Participants answered using a seven-point Likert scale ranging from ``Complicate'' (point -3) to ``Help'' (point 3).
The distribution of social valence data is plotted in Figure~\ref{fig:socialValence}.

Shapiro-Wilk tests reported a non-normal distribution regarding social valence values.
Therefore, a Wilcoxon Signed-Rank Test was conducted to analyse the differences between both conditions.
% Nevertheless, players of the Mutual Help version showed a stronger tendency to ``Help" the other player compared to ``Complicate" their gameplay.

\begin{figure}[!htbp]
	\centering
	\includegraphics[width=0.95\columnwidth]{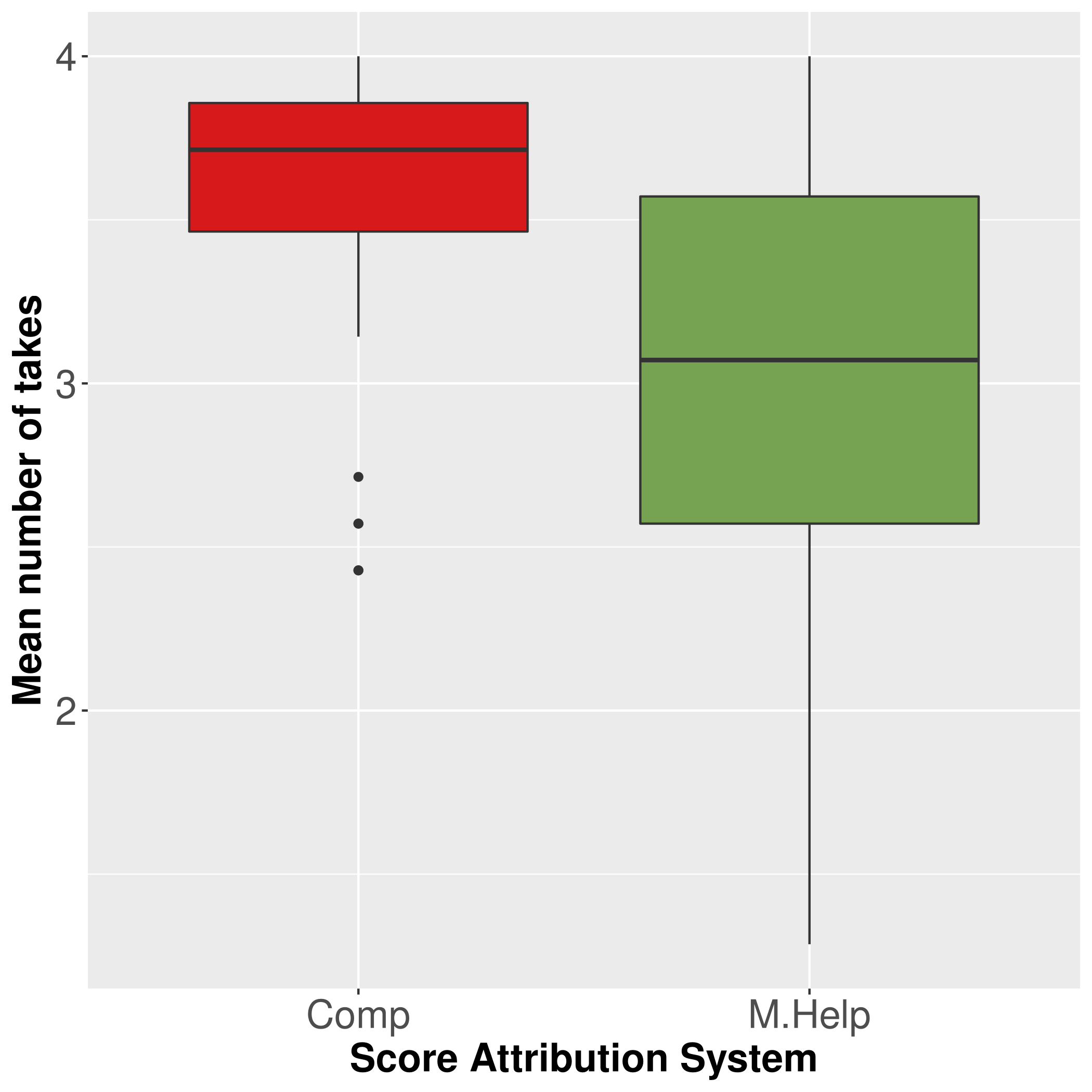}
	\caption{Distribution of average number of \textit{take} actions by score system.}
	\label{fig:takes}
\end{figure}

\begin{figure}[!htbp]
	\centering
	\includegraphics[width=0.95\columnwidth]{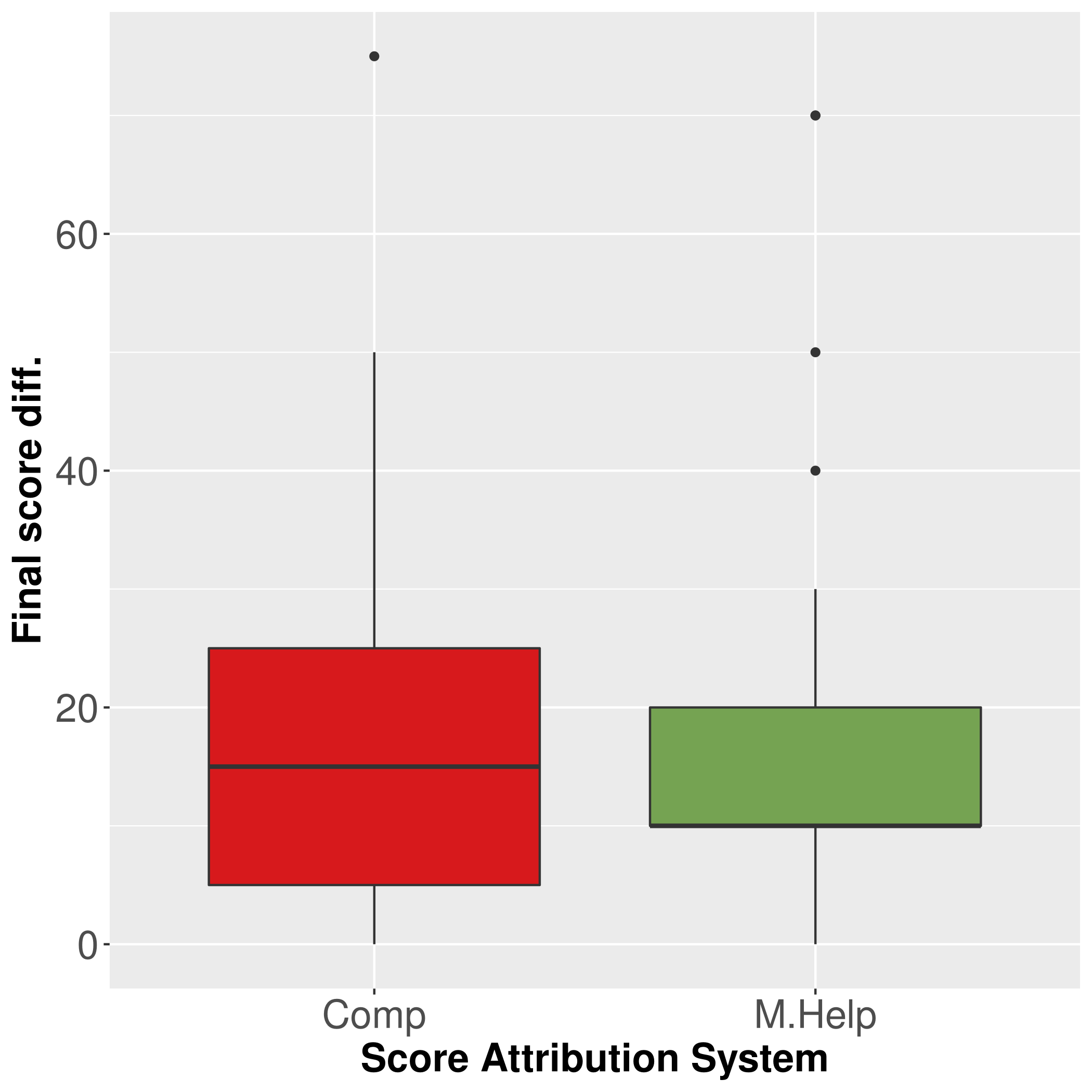}
	\caption{Difference between final scores by score system.}
	\label{fig:finalScores}
\end{figure}

Results of applying the Wilcoxon test indicated large-sized significant differences in the social valence of interactions between our game versions \((Z \approx 6.60, p < .001, r \approx .812)\).
In particular, the data has shown a tendency for the \textit{Competition} version to drive participants towards ``Complicate'' \((M=-1.5; Mdn = -2;SD\approx1.60)\), and a tendency for the \textit{Mutual Help} version to drive participants towards ``Help'' \((M\approx0.74;Mdn = 1; SD\approx1.64)\).
% \(T = 1785.50, p < .001, r = -.79\).
These tendencies deem us to conclude that indeed \textit{reward has a large effect on the social valence of interactions in a two-player scenario}.
% Additionally, there was a large effect size.
Furthermore, these findings shed light on how a reward-based system aimed at fostering competition should be modeled.
By rewarding players and punishing their peers when the interaction scenario implied the attainment of a letter both players needed, participants implicitly perceived that situation as a challenge, in which they had to focus more on competing.
As for our approach to induce positive valence, the interquartile range was not as close to the intended extreme.
We believe that this less pronounced tendency to ``Help" was due to a wider range of possible interpretations, which depend on what strategies the players applied.
% as in the \textit{Competition} version.
% {\color{red}
% In particular, the number of \textit{take} actions was considerably higher \((Mdn = 3.07)\) compared to the number of gives \((Mdn = .93)\).
% }
Another hypothesis is that, as it is more common in ludic culture, players focused more on completing their task rather than interacting with their peers.
Nonetheless, as seen in Figure~\ref{fig:takes}, there is a significant difference in the distribution of \textit{take} actions between versions, \((Z \approx 5.53, p < .001, r=.680)\), with an expected tendency for the \textit{Competition} version to drive players to take more letters \((M \approx 3.60, Mdn \approx 3.71, SD \approx 3.77)\) than players of the \textit{Mutual Help} version \((M\approx2.99, Mdn\approx3.07, SD\approx.715)\).
In addition, we did not find a statistically significant difference in the final score between peers for the \textit{Competition} \((M=20, Mdn = 15, SD\approx17.28)\) and \textit{Mutual Help} \((M\approx18.5, Mdn = 10, SD\approx16.98)\) versions, \(Z \approx 1.10, p \approx 0.27\).
However, as can be observed in Figure~\ref{fig:finalScores}, the \textit{Competition} version led to a larger variance in differences in scores compared to the \textit{Mutual Help} system, which is expected due to the natures of the two promoted behaviors.
% This effect is in line with the fostered behavior for each scoring system.
Players in the \textit{Competition} version were expected to increase the difference between scores, due to the punishing imposed when a player took a needed shared letter.
Conversely, \textit{Mutual Help} players were expected to decrease the difference between scores, as the score system rewarded players for aiding the other player's task progression.
Nevertheless, we expected to find significant differences between the two conditions  due to the way rewards were computed for each case.
For instance, contrary to the \textit{Mutual Help} version, rewards in the \textit{Competition} setting did not keep a fixed delta between players.
While the scoring system would promote a delta of ten if both words needed the same letter, the other two meaningful \textit{take} actions only led to a delta of five.
These differences may have hindered how much players chose the optimal strategy of progressing in the task and spoil the other player's progression at the same time.
Moreover, we believed that the difference between scores would be close-to-zero in the \textit{Mutual Help} version.
We argue that this effect may be a result of either the complexity of the reward function, which takes into account task progression, or the competitive setting.
Notably, the optimal strategy may not have been fully understood by players at the finer granularity of task progression, although they showed a predisposition to keep a balance between scores.
Altogether, the observed results lead us to believe that we were able to address both research questions.

Although we found strong results, there are some limitations worth discussing.
Our experiment was conducted with both players \textit{in loco}, contrary to other research such as Vegt et al.~\cite{Vegt2016} who used no direct contact between participants.
This may have lead players to feel social pressure regarding how they interacted with their peer, since their reaction could be communicated to the other player.
Nevertheless, we aimed at diminishing this effect by recruiting participants in pairs, preferably knowing each other.
In addition, we limited the number of actions a player could perform per level at four.
On the one hand, this decision transmits the cost of opportunity to players as they could not waste actions in meaningless interactions.
On the other hand, it may have subverted specific strategies that require more actions to show effects.
Moreover, the limited number of actions may also create a bias, limiting more ``altruistic" interactions.
Finally, we set the number of common letters in a word at two so that players had fixed opportunities to interact, instead of allowing the option to have totally independent gameplays.
We believe that future work should address these limitations in order to verify whether our results are not strongly affected by these variables.

\section{Conclusions and Future Work}
\label{sec:conclusion}

The aim of our work was to investigate whether rewards influence the social valence of interdependent interactions. % in a two-player setting.
To achieve that goal, we deployed different reward attribution systems in a two-player word-matching game called Message Across. In particular, we considered two score attribution systems aimed at moderating two interdependent interaction styles with opposite social valence: \textit{Competition} and \textit{Mutual Help}. We then measured what effects these systems had in players strategies and reported interaction intentions. The results indicated that players significantly embraced different strategies and reported different intentions when playing each version, aligned with our modeled valence poles.
The results presented in this work contribute to advancing the state-of-the-art of knowledge in several aspects.
First, we tested and validated that strategies aimed for fostering competition should include score functions that reward and punish players based on how they interact in a game.
In particular, competitive score systems should aim at increasing the score difference.
As we have shown, this type of reward functions lead players to change their social valence while interacting with their peers.
Secondly, mutual help may be promoted using score functions that try to bridge the gap between the scores of both players.
While still rewarding players for fulfilling their task, interactions that help peers should also be considered as valid as the former.
Finally, another interesting point is how participants were able to understand the different score systems as we designed them without being informed of those details.
In the light of this, our results may be applied as game design guidelines to create reward functions aiming at fostering interaction behaviors with distinct valence values.

Future work includes investigating whether individual differences such as personality had an effect on how people varied their social valence values.
For example, we believe that people with higher Agreeableness~\cite{mccrae1992introduction} would be more prone to ``Help" other players compared with people with low Agreeableness.
We also believe that studying different sized words or words with different numbers of shared letters is worthwhile to check how the length and type of the task may have an impact in the players' strategies and perceptions.
In addition, we would like to study how much participants adopted the optimal strategy while playing both versions.
Measuring more in-depth the flow of the game would allow us to study at a finer granularity whether the score differences in distinct versions were a result of the proficiency of the participants.
Finally, we can study our social valence scale more in-depth.
For instance, as we mentioned in Section~\ref{sec:method}, cooperation between peers can depend on the focus of the interaction, i.e. the focus of the players can vary between themself and others.
Therefore, we can create more reward-based systems to promote interaction styles such as \textit{Extreme Altruism}.
In the same light, competition can be studied in-depth to investigate whether players interact following an healthy competition, where they try to fulfill their task while competing with their peers, or a toxic competition, where they do not care for their task or score because their main objective is to complicate the other's gameplay.

% Our findings provide new contributions to the field of training and educating people, which has been a focus of multiplayer serious games recently.
% As we mentioned, our results support that rewards may be used as a means to promote interaction styles varying on social valence.
% Systems such as GIMME~\cite{gomes2019gimme} that simulate groups of people interacting with one another may leverage how scores affect these interactions to empower collective teaching in multiplayer settings.
% In particular, promoting interaction styles with rewards allows researchers to take a more human approach to the integration of agents that simulate people in serious games, besides adding expressiveness to their simulation models.
Our findings provide new contributions to the field of automatic education and training, due to the importance of behavior promotion for this research topic.
Models such as GIMME~\cite{gomes2019gimme}, that aim to optimize the collective ability of groups of people interacting with one another, may use scores to mediate students' interactions, thus empowering collective teaching in multiplayer settings.
Additionally, the moderation of interactions using rewards allows researchers to take a more human approach to the integration of agents simulating serious games players, besides adding expressiveness to their simulation models.

%\section*{Acknowledgment}

\bibliographystyle{IEEEtran}
\bibliography{sample}

\end{document}